\documentclass[twocolumn,showpacs,preprintnumbers,amsmath,amssymb,floatfix]{revtex4}
\usepackage{graphicx}

\begin{document}

\title{Electronic coherence in metals:
comparing weak localization and time-dependent conductance fluctuations}

\author{A. Trionfi, S. Lee, and D. Natelson}

\affiliation{Department of Physics and Astronomy, Rice University, 6100 Main St., Houston, TX 77005}

\date{\today}


\begin{abstract}
Quantum corrections to the conductivity allow experimental assessment
of electronic coherence in metals.  We consider whether independent
measurements of different corrections are quantitatively consistent,
particularly in systems with spin-orbit or magnetic impurity
scattering.  We report weak localization and time-dependent universal
conductance fluctuation data in quasi-one- and two-dimensional AuPd
wires between 2~K and 20~K.  The data inferred from both methods are
in excellent quantitative agreement, implying that precisely the same
coherence length is relevant to both corrections.
\end{abstract}

\maketitle

Quantum coherence of electrons in solids remains a topic of much
interest.  Technologically, coherent control and manipulation of
electrons is relevant in proposed novel
devices\cite{LiangetAl00Nature,CapassoetAl96JMP}.  Scientifically, the
mechanisms and temperature dependence of decoherence are of
fundamental importance\cite{Imry97book}, and have profound implications
for the ground state of metals in the presence of disorder.  Quantum
corrections to the conductivity allow coherence to be
examined experimentally.  Specific corrections that have been used
include the weak localization (WL)
magnetoresistance\cite{Bergmann84PR}, universal conductance
fluctuations as a function of magnetic
field\cite{SkocpoletAl86PRL,WashburnetAl92RPP} (MFUCF), time-dependent
universal conductance fluctuations
(TDUCF)\cite{BirgeetAl89PRL,Giordanobook}, and Aharonov-Bohm
oscillations\cite{WebbetAl85PRL}.

These corrections result from interference between electronic
trajectories on length scales shorter than the coherence length,
$L_{\phi}\equiv \sqrt{D \tau_{\phi}}$, where $D$ is the electron
diffusion constant and $\tau_{\phi}$ is the timescale over which the
phase of the electron's wave function is perturbed strongly by
environmental degrees of freedom.  It is interesting to ask whether
precisely the same time (length) scales are relevant to the various
quantum corrections.  For example, the electron ``out-scattering''
time (for scattering out of a particular momentum state) in the
Boltzmann equation with screened Coulomb interactions has a different
temperature dependence\cite{AbrahamsetAl81PRB} than the coherence time
for weak localization\cite{AltshuleretAl82,Imry97book}, and has been
suggested as relevant to UCF\cite{Blanter96PRB}.  One must also
consider whether other complications ({\it e.g.}  spin-orbit coupling;
scattering from dilute magnetic impurities) affect the inferred values
of $L_{\phi}$ identically.  Subtleties are known to exist regarding
magnetic impurities in Aharonov-Bohm rings\cite{SternetAl90PRA}.
These questions have particular relevance as recent publications
concerning saturation\cite{MohantyetAl97PRL} of $L_{\phi}^{\rm WL}(T)$
as $T\rightarrow 0$ have included comparisons with Aharonov-Bohm
experiments\cite{PierreetAl02PRL} and MFUCF
data\cite{MohantyetAl03PRL}.

Weak localization results from electron trajectories that form closed
loops, and their time-reversed conjugates.  With no spin-orbit
scattering and at zero magnetic field, such pairs constructively
interfere, leading to a lowered conductance.  Strong spin-orbit
interactions lead instead to destructive interference, and a
conductance increase at zero magnetic field.  Magnetic flux through
such a loop suppresses these interference effects,
resulting in a magnetoresistance with a field scale that reflects
$L_{\phi}^{\rm WL}$ and the sample geometry.

Time-dependent UCF result from changes in defects' positions that
alter the phases of interfering trajectories, and hence the
conductance within a coherent volume.  With an appropriate broad
distribution of defect relaxation times, the resulting noise power has
a $1/f$ dependence\cite{FengetAl86PRL}.  Applied magnetic flux
suppresses the cooperon contribution to the
fluctuations\cite{Stone89PRB} over a field scale related to
$L_{\phi}^{\rm TDUCF}$, reducing the noise power by a factor of two.
As $T\rightarrow 0$, $L_{\phi}^{\rm TDUCF}$ grows relative to sample
size, $L$, and thermal smearing is reduced, leading to an increase of
TDUCF noise power.  For WL and the field dependence of
TDUCF\cite{LeeetAl87PRB,Stone89PRB,MoonetAl97PRB}, the quasi-1D limit
occurs in samples of width $w$ and thickness $t$ when $w,t <
L_{\phi}$, while the quasi-2D limit occurs when $t < L_{\phi} < w$.
The thermal length, $L_{T}$, is defined as $L_{T} \equiv \sqrt{\hbar
D/k_{\rm B}T}$ and is important for determining the magnitude of UCF.

\begin{table}
\caption{Samples used in magnetotransport and noise measurements.  
Free electron density of states for Au used to calculate $D$: $1 \times 10^{47}$~m$^{-3}$J$^{-1}$, from Ref.~\protect{\cite{Ashcroft75}}.  Sample D was
deliberately contaminated with additional ferromagnetic impurities.}
\begin{tabular}{c c c c c}
\hline \hline
Sample & $w$~[nm] & $t$~[nm] & $R_{\Box}$~[$\Omega$] & $D$ [m$^{2}$/s]  \\
\hline
A & 43 & 9 & 32.1 & 1.34 $\times 10^{-3}$ \\
B & 35 & 9 & 31.5 & 1.34 $\times 10^{-3}$ \\
C & 500 & 6.5 & 84.5 & 7.9 $\times 10^{-4}$ \\
D & 500 & 8.5 & 47.9 & 9.6 $\times 10^{-4}$ \\
\hline
\hline
\end{tabular}
\label{tab:samples}
\end{table}

Previous experimental comparisons between $L_{\phi}^{\rm WL}$ and
$L_{\phi}^{\rm TDUCF}$ were equivocal.  In quasi-2d silver
films\cite{McConvilleetAl93PRB,HoadleyetAl99PRB}, the two lengths
agreed quantitatively only above a temperature where $L_{\phi}^{\rm
TDUCF} \approx L_{\rm SO}$, the spin-orbit scattering length.  At 2~K,
$L_{\phi}^{\rm WL}\sim 2 \times L_{\phi}^{\rm TDUCF}$.  The results
were interpreted as consistent with $L_{\phi}^{\rm WL}$ set by Nyquist
scattering and $L_{\phi}^{\rm UCF}$ determined by the out-scattering
rate\cite{Blanter96PRB}.  Similar investigations in quasi-1d Li
wires\cite{MoonetAl97PRB} showed better agreement between
$L_{\phi}^{\rm WL}$ and $L_{\phi}^{\rm TDUCF}$ in a weak spin-orbit
system, but data were limited.  A theoretical
reexamination\cite{AleineretAl02PRB} now predicts agreement between
these lengths in both quasi-1d and quasi-2d systems when decoherence
arises from electron-electron interactions.  This agreement is
expected to remain true\cite{Aleinerpc} as long as other decoherence
mechanisms ({\it e.g.} electron-phonon; spin-flip scattering) do not
involve small ($<< k_{\rm B}T$) energy
transfers\cite{AleineretAl02PRB}.

We compare $L_{\phi}^{\rm WL}$ and $L_{\phi}^{\rm
TDUCF}$ in mesoscopic AuPd wires in both the quasi-1 and 2D limits.
The AuPd is known to have extremely strong spin-orbit
scattering\cite{LinetAl87PRB}.  One quasi-2d sample was
deliberately contaminated with ferromagnetic impurities.  We find that
coherence lengths inferred from both WL and TDUCF are in strong
numerical agreement between 2 and 20 K, independent of dimensionality
and magnetic impurity concentration.  This agreement implies that
coherence lengths inferred from these different experimental
techniques may be compared {\em quantitatively}, even in the presence
of significant spin-orbit interactions and decoherence due to
spin-flip scattering.

\begin{figure}[h!]
\begin{center}
\includegraphics[clip, width=7.5cm]{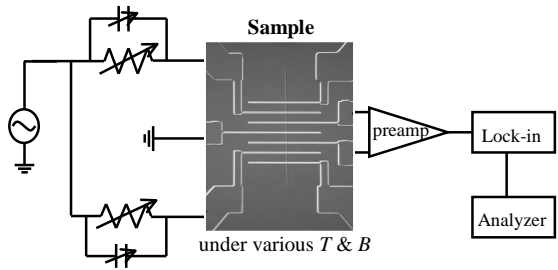}
\end{center}
\vspace{-3mm}
\caption{Noise measurement scheme.  Trimming capacitors are used to null away any capacitive phase differences between the two bridge halves.  The samples consist of 7 leads with only five consecutive leads used. }
\label{fig:scheme}
\end{figure}

All samples were fabricated by electron beam lithography on undoped
GaAs substrates.  Figure~\ref{fig:scheme} shows the sample
configuration, and the parameters for each sample are described in
Table~\ref{tab:samples}.  For each sample between 6.5 and 9~nm of
Au$_{0.6}$Pd$_{0.4}$ was evaporated to create the wire, followed by a
second lithography step to create the leads.  The leads consisted of
1.5~nm thick Ti and followed by 25~nm of Au.  Each segment of wire
between Ti/Au leads was 10~$\mu$m in length, and each wire consisted
of seven segments.  All evaporations were performed via an electron
beam evaporator at $\sim 5\times 10^{-7}$~mB.  Knowing the purity of
the starting material, the AuPd alloy likely contains magnetic
impurities at the few parts per million level, as discussed below.  To
produce a sample (D) with a higher magnetic impurity concentration,
roughly 2.5~nm of Ni$_{0.8}$Fe$_{0.2}$ was evaporated with the sample
shutter {\it closed} immediately prior to AuPd deposition.  Contact
resistances were less than 30~$\Omega$.  Diffusion constants were
calculated using the Einstein relation and the density of states for
bulk Au\cite{Ashcroft75}.

Samples were measured in a $^{4}$He cryostat and initially
characterized by four-terminal resistance versus temperature in a 3~T
magnetic field normal to the wire.  Magnetic impurity concentrations
in all samples were sufficiently low that no Kondo upturn in
resistivity was distinguishable.  Currents from 10~nA to
10~$\mu$A were set at each temperature such that no Joule heating was
detected in $R(T)$.

All noise measurements were performed using a five-terminal ac bridge
technique\cite{Scofield87RSI} with a carrier frequency of 600~Hz.  No
drive current dependence was observed in either WL or TDUCF until
currents large enough to affect $R(T)$.  WL magnetoresistance was
measured in a four-terminal configuration while varying a
perpendicular magnetic field between $\pm 1.25$~T.  For the TDUCF, the
demodulated lock-in output was fed into a dual channel signal analyzer
to transform the signal into the frequency domain.  A typical
frequency range was 78~mHz to 1.5~Hz.  Background pre-amp noise was
measured simultaneously using the out-of-phase output of the lock-in,
and subtracted from the in-phase noise signal.  Excellent agreement
with a $1/f$ dependence of the noise power was found consistently.  As
expected for TDUCF, the measured noise power increased as $T
\rightarrow 0$, and depended nontrivially on $B$ as described below.

\begin{figure}[h!]
\begin{center}
\includegraphics[clip, width=7.5cm]{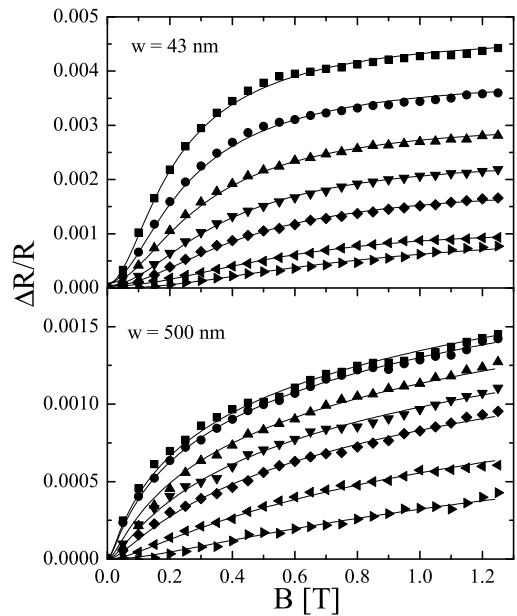}
\end{center}
\vspace{-5mm}
\caption{Weak (anti)localization magnetoresistance at various temperatures for a 43~nm wide wire (quasi-1D, top) and a 500~nm wide wire (quasi-2D, bottom), with fits to Eqs.~(\ref{eq:1dwl},\ref{eq:2dwl}), respectively.  Top to bottom, temperatures are 2~K, 4~K, 6~K, 8~K, 10~K, 14~K, and 20~K. }
\label{fig:wlmr}
\vspace{-2mm}
\end{figure}

Figure~\ref{fig:wlmr} shows typical magnetoresistance curves for a
quasi-1D and a quasi-2D sample.  The WL magnetoresistance
formulae with strong spin-orbit interactions for 1D and 2D are
\begin{equation}
\frac{\Delta R}{R}\big{\vert}_{\rm 1d} = - \frac{e^{2}}{2 \pi \hbar}\frac{R}{L}\left[\frac{1}{L_{\phi}^{2}}+\frac{1}{12}\left(\frac{w}{L_{B}^{2}}\right)^{2}\right]^{-1/2}
\label{eq:1dwl}
\end{equation}
\begin{equation}
\frac{\Delta R}{R}\big{\vert}_{\rm 2d} = \frac{e^{2}}{4 \pi^{2} \hbar}R_{\Box}\left[\psi\left(\frac{1}{2}+\frac{1}{2}\frac{L_{B}^{2}}{L_{\phi}^{2}}
\right)-\ln\left(\frac{1}{2}\frac{L_{B}^{2}}{L_{\phi}^{2}}\right)\right]
\label{eq:2dwl}
\end{equation}
respectively\cite{PierreetAl03PRB,LinetAl87PRB}.  
The differing forms result from
divergences that depend on dimensionality\cite{Altshulerbook}.  Note
that $\Delta R = R(B)-R(B=\infty)$ for Eq.~(\ref{eq:1dwl}) while
$\Delta R = R(B)-R(B=0)$ for Eq.~(\ref{eq:2dwl}).  Here $\psi$ is the
digamma function, $L_{B}$ is the magnetic length and is defined as
$L_{B}\equiv \sqrt{ \hbar/2eB}$ , and $R_{\Box}$ is the sheet
resistance.  In fitting the quasi-1d magnetoresistance data, at 2~K
the width $w$ was allowed to vary, and was then fixed for all other
fits.  Widths found in this manner (43~nm and 35~nm) were consistent
with both electron micrographs and estimates based on measured
resistances and $R_{\Box}$ found in codeposited films.  Including
$L_{\rm SO}$ as a fit parameter leads to $L_{\rm SO} \lesssim$~10~nm,
with little impact on $L_{\phi}$.

\begin{figure}[h!]
\begin{center}
\includegraphics[clip, width=7.5cm]{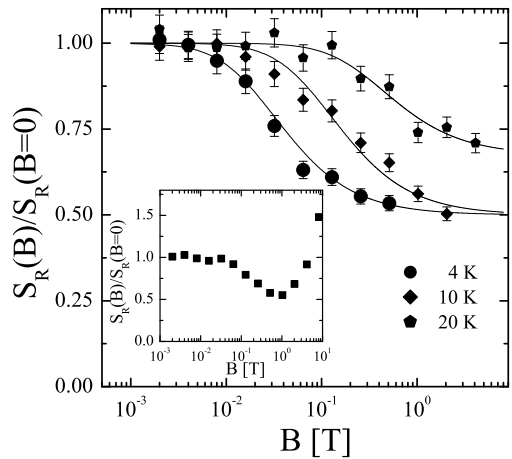}
\end{center}
\vspace{-5mm}
\caption{Normalized noise power as a function of magnetic field for a 500~nm wide wire.  The 20~K point does not drop by a full factor of 2 due to local interference noise.  Inset: At high field there is a large upturn in the noise for a sample deliberately dosed with additional magnetic impurities (500~nm wide sample at 4~K).}
\label{fig:noisefield}
\vspace{-5mm}
\end{figure}

Figure~\ref{fig:noisefield} shows examples of the normalized noise
power $(S_{R}(B)/S_{R}(B=0))$ as a function of field.  As in WL, the
characteristic field scale involves magnetic flux through loop-like
trajectories, with a lower field corresponding to a larger
$L_{\phi}^{\rm TDUCF}$.  The normalized noise power as a function of
$B$ is the crossover function, $\nu(B)$, and depends on
dimensionality.  Analytical expressions for the crossover functions in
the strong spin-orbit limit have recently been
calculated\cite{Aleinerpc}:
\begin{equation}
\nu_{\rm 1d}(B)=1- \frac{x}{2}\left(\frac{Ai(x)}{Ai'(x)}\right)^{2}
\label{eq:1dnoise}
\end{equation}
where $x \equiv L_{\phi}^{2}/(3(\hbar/B e w)^{2})$,
and
\begin{equation}
\nu_{\rm 2d}(B)=\frac{1}{2}+ \frac{L_{B}^{2}}{4 L_{\phi}^{2}}\psi'\left(\frac{1}{2}+\frac{L_{B}^{2}}{2 L_{\phi}^{2}}\right),
\label{eq:2dnoise}
\end{equation}
respectively.  These functional forms are strictly valid when
$\hbar/\tau_{\phi}<< k_{\rm B}T$.  Here $Ai(x)$ is the Airy
function, and $\psi'(x)$ is the derivative of the digamma function.

Previous $L_{\phi}^{\rm TDUCF}$
extractions\cite{McConvilleetAl93PRB,MoonetAl97PRB,HoadleyetAl99PRB}
have used the numerical crossover function calculated by
Stone\cite{Stone89PRB} for quasi-2d samples, as well as an approximate
analytical form derived by Beenakker and van
Houten\cite{BeenakkeretAl88PRB} for 1D samples.  Comparisons between
the analytic and numerical forms demonstrated $T$-independent
differences (numerical $>$ analytical) of roughly 14\% for quasi-1D
samples and 3\% for quasi-2D samples.  As in the WL data, the
resulting fits were essentially unaffected by including $L_{\rm SO}$
as a fit parameter, since $L_{\rm SO}$ is so short.

To account for field-independent local interference
noise\cite{PelzetAl87PRB,Hershfield88PRB} at higher temperatures, a
second fitting parameter, $z$, the fraction of noise that is due to
UCF, was introduced into the fitting function, $f(B)=(1-z) + z\nu(B)$.
We found that $z$ was indistinguishable from 1 for all temperatures
measured except for 20~K in the 2d samples, when $z \approx 0.68$.
All fits and confidence intervals were determined by nonlinear
$\chi^{2}$ minimization and analysis.

The inset to Fig.~\ref{fig:noisefield} shows the normalized noise
power as a function of field for the magnetically contaminated sample.
The upturn at large fields is a suppression of spin-flip decoherence
as the Zeeman splitting of the magnetic impurities exceeds $k_{\rm
B}T$.  An analogous upturn has been observed in investigations of Li
wires\cite{MoonetAl97PRB} and in recent Aharonov-Bohm measurements in
Cu rings\cite{PierreetAl02PRL}.  Some upturn is visible at the highest
$B/T$ ratio in {\it all} of our samples, consistent with some magnetic
impurities even in nominally ``clean'' devices.  Note that the effects
of spin-flips on WL and TDUCF depend on the ratio of the spin-flip
time and the temperature-dependent impurity Korringa
time\cite{PierreetAl03PRB}.  For $T> \sim 40~{\rm mK}\times$ the ppm
concentration of magnetic impurities, spin-flip scattering should
involve large energy transfers\cite{Aleinerpc}, and affect WL and
TDUCF identically.  For our samples (with $\sim$ a few ppm
impurities), this crossover is well below 1~K, outside the regime of
these experiments.

\begin{figure}[h!]
\begin{center}
\includegraphics[clip, width=7.5cm]{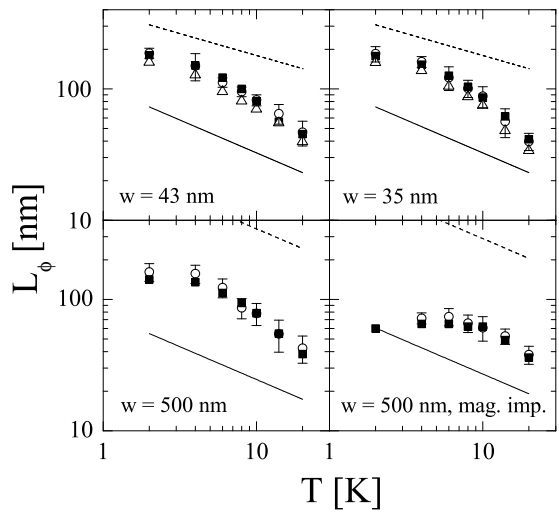}
\end{center}
\vspace{-5mm}
\caption{\small The coherence lengths of the four samples, as
indicated.  Solid squares: WL data; open circles: Beenakker/van
Houton/Stone\protect{\cite{BeenakkeretAl88PRB,Stone89PRB}} fit to
TDUCF field dependence; open triangles: Aleiner fit
(Eq.~(\ref{eq:1dnoise})) to TDUCF field dependence.  Only one TDUCF
fit is shown for the 2D samples since both fits result in the same
number to within 3\%.  Dashed lines are predicted values for $L_{\phi}$
assuming decoherence is dominated by Nyquist scattering\protect{\cite{AleineretAl99WRM}} and using 
sample parameters from Table~\ref{tab:samples}.  Solid lines are $L_{T}$ values
calculated from the same sample parameters.}
\label{fig:LphiT}
\end{figure}

The resulting coherence lengths from both WL and UCF measurements are
shown in Figure~\ref{fig:LphiT}.  The temperature dependence becomes
steeper as electron-phonon scattering increases.  Clearly, over the
temperature range measured the coherence lengths inferred from the two
techniques are in excellent agreement.  This agreement remains strong
even in the presence of magnetic impurity scattering significant
enough to suppress the coherence length by more than a factor of two.
This strongly supports the theoretical
statement\cite{AleineretAl02PRB} that weak localization and UCF
measurements probe {\it precisely} the same coherence physics, even in
the presence of strong spin-orbit and magnetic impurity scattering.

The agreement is noteworthy.  First, $L_{\phi}$ values at the lowest
temperatures are below those predicted from the pure Nyquist
electron-electron dephasing (for example, see
Ref.~\cite{AleineretAl99WRM}).  This is not surprising given the
presence of magnetic impurities in the AuPd, as described above.
Second, the agreement persists even though $\hbar/\tau_{\phi}$ is
never $<< k_{\rm B}T$, suggesting that
Eqs.~(\ref{eq:1dnoise},\ref{eq:2dnoise}) are robust even when that
constraint is somewhat relaxed.

These results leave open the question of why the coherence lengths in
Ag inferred from WL and TDUCF have differing temperature dependences\cite{McConvilleetAl93PRB,HoadleyetAl99PRB}.
The simplest explanation would involve some subtle effect from triplet
channel interactions that is only relevant when $L_{\rm SO} \sim
L_{\phi}$.  Until further theoretical and experimental investigations
address this regime, any quantitative attempts to compare different
coherence phenomena in materials with intermediate spin-orbit
scattering should be done with care.

We have carefully measured weak localization magnetoresistance and the
magnetic field dependence of time-dependent universal conductance
fluctations in mesoscopic AuPd wires.  By comparing the coherence
lengths inferred from these data, we have shown that $L_{\phi}^{\rm
WL}$ and $L_{\phi}^{\rm TDUCF}$ are in quantitative agreement, even in
the presence of potentially subtle effects such as strong spin-orbit
scattering and spin-flip contributions to dephasing .  Numerical
consistency should therefore be expected between complementary UCF and
WL measurements of electronic coherence.

We would like to thank N.O. Birge for his helpful advice concerning
noise measurements, and I.L. Aleiner and A.D. Stone for discussions of
the theory.  This work was supported by DOE grant
DE-FG03-01ER45946/A001.




\end{document}